\documentclass[12pt]{spieman}  
\usepackage{amsmath,amsfonts,amssymb}
\usepackage{graphicx}
\usepackage{setspace}
\usepackage{tocloft}
\usepackage{lineno}
\usepackage{xcolor}

\title{Dust stripping in cluster galaxies: a PRIMA perspective}

\author[a,b]{Alessandro Boselli}
\author[c]{Marc Sauvage}
\author[a]{Laure Ciesla}

\affil[a]{Aix Marseille Universit\'e, CNRS, CNES, LAM, Marseille, France}
\affil[b]{INAF - Osservatorio Astronimico di Cagliari, Via della Scienza 5, 09047 Selargius (CA), Italy}
\affil[c]{AIM, CEA, CNRS, Universit\'e Paris-Saclay, Universit\'e Paris Diderot, Universit\'e Paris Cit\'e, F-91191 Gif-sur-Yvette, France}

\cftpagenumbersoff{figure}
\cftpagenumbersoff{table} 

\begin{document} 
\maketitle

\begin{abstract}
The evolution of galaxies in rich environments such as clusters and groups can be significantly perturbed during their interaction with nearby companions (tidal interactions) or with the hot 
intracluster medium (ICM) trapped within the gravitational potential well of the massive host halo (ram pressure stripping). Both gravitational perturbations and the external pressure exerted 
by the hot ICM on the galaxy ISM during its high velocity journey within the cluster are able to remove most, if not all of it, producing extended tails of stripped material. Along with the 
different gas phases (cold atomic and molecular, ionised, hot), these perturbations can remove also dust, thus contributing to the pollution of the ICM. Probe Infrared Mission for Astrophysics 
(PRIMA) is offering a unique opportunity to observe this dust component, a crucial ingredient in the energetic balance of the stripped gas. We analyse how the two instruments onboard of PRIMA, 
PRIMAger (in imaging and polarimetric mode) and far-IR enhanced survey spectrometer (FIRESS), can be used to observe a selected sample of ram pressure stripped tails detected at other frequencies 
(HI, CO, Halpha, X-rays). These data can be used to determine the relative distribution of the dust component with respect to that of the other gas phases, derive its temperature, calculate 
different gas physical parameters (electron density, photoelectric heating efficiency, gas metallicity), and quantify the strength of the turbulent magnetic fields, all fundamental parameters 
used to constrain the most recent hydrodynamic simulations of gas stripping in clusters. The ultimate aim of this research is that of understanding the fate of the cold stripped material once 
mixed with the surrounding hot medium and study under which condition it can collapse into giant molecular clouds (GMC) to form new stars.  
\end{abstract}


{\noindent \footnotesize\textcolor{red}{*}Alessandro Boselli \linkable{alessandro.boselli@lam.fr} }


\begin{spacing}{1}   

\section{Introduction}
\label{sect:intro}  
Galaxies inhabiting rich environments have physical properties significantly different than those in low-density regions. They are principally relaxed, pressure-supported systems 
(ellipticals, lenticulars), composed of evolved stars, with a very reduced amount of gas and dust, while their counterparts in the field are generally gas-rich, star forming rotating 
systems ([\citenum{Dressler1980}, \citenum{Whitmore1993}, \citenum{Dressler1997}, \citenum{Postman2005}, \citenum{Gavazzi2010}]). Furthermore, star-forming galaxies in high-density regions 
are also different from those located in low-density environments. They are generally gas-poor, in particular in their atomic gas phase (e.g. [\citenum{Haynes1984}, \citenum{Cayatte1990}, 
\citenum{Solanes2001}, \citenum{Gavazzi2005}, \citenum{Catinella2013}]). They also have a reduced molecular gas content ([\citenum{Fumagalli2009}, \citenum{Boselli2014c}, \citenum{Zabel2019}, 
\citenum{Zabel2022}]). Because of this observed lack of gas, late-type systems in high-density regions have also a reduced star formation activity with respect to their counterparts in the 
field (e.g. [\citenum{Kennicutt1983}, \citenum{Gavazzi1998}, \citenum{Lewis2002}, \citenum{Gomez2003}, \citenum{Boselli2014b}, \citenum{Boselli2016b}]).  

All this evidence clearly indicates that the environment in which they reside plays a major role in shaping galaxy evolution (environmental quenching, [\citenum{Peng2010}]). Different physical 
mechanisms have been proposed in the literature to explain these observed differences. They can be broadly divided into two major families (e.g. [\citenum{Boselli2006a}, \citenum{Boselli2014c}]): 
the gravitational interactions between the different group/cluster members (e.g. [\citenum{Merritt1983},\citenum{Moore1998}]) or with the gravitational potential well of the cluster itself 
(e.g. [\citenum{Byrd1990}]), and the hydrodynamic interactions between the cold interstellar medium (ISM) of the galaxies and the hot intracluster medium (ICM) trapped within the gravitational 
potential well of the high-density region. These include ram pressure stripping ([\citenum{Gunn1972}]), viscous stripping ([\citenum{Nulsen1982}]), thermal evaporation ([\citenum{Cowie1977}]), and starvation ([\citenum{Larson1980}]). Among these mechanisms, the one dominant in massive 
($M_{halo}$ $\gtrsim$ 10$^{14}$ M$_{\odot}$) local clusters seems to be the ram pressure 
stripping exerted by the hot ICM on the cold ISM of galaxies moving at high velocity ($\sim$ 1000 km s$^{-1}$) within it ([\citenum{Boselli2022}]). 
This mechanism is able to remove the gas of the ISM in all its phases, principally the one located in the outer disc of the perturbed galaxies, forming extended tails of stripped 
material now commonly observed at different frequencies in several local ([\citenum{Gavazzi2001}, \citenum{Chung2007}, \citenum{Yagi2010}, \citenum{Poggianti2017}, \citenum{Roberts2021a}, 
\citenum{Roberts2021b}]) and intermediate-redshift (e.g. [\citenum{Yagi2015}, \citenum{Boselli2019}]) clusters.

Along with the different gas phases of the ISM (cold atomic, molecular, ionised, hot), these perturbations are expected to remove also the interstellar dust, which in normal galaxies is 
generally well mixed with the gaseous component (e.g. [\citenum{Burstein1982}, \citenum{Remy-Ruyer2014}, \citenum{Whittet2022}]). Direct observation of dust in the stripped tails of perturbed 
galaxies, however, is still lacking principally due to the limited sensitivity of the instruments used in the observations. Using far-infrared Herschel data gathered during the Herschel 
Reference Survey (HRS, [\citenum{Boselli2010}]) and the Herschel Virgo Cluster Survey (HeViCS, [\citenum{Davies2010}]), [\citenum{Longobardi2020b}] observed an asymmetric distribution of 
the cold dust component in a few galaxies in the Virgo cluster showing an extended tail of gas witnessing an ongoing ram pressure stripping event. [\citenum{Kenney2015}], using high-resolution 
Hubble Space Telescope (HST) imaging data, discovered prominent filaments of dust escaping from the disc of NGC 4921, a massive spiral in the Coma cluster, also suggesting that dust is removed 
along with the different gaseous components. There is also evidence that several stripped objects in local clusters have dust truncated discs ([\citenum{Cortese2010}]) as those observed in HI 
(e.g. [\citenum{Cayatte1994}]) and CO ([\citenum{Fumagalli2009}, \citenum{Boselli2014a}]). The most representative case is probably the massive spiral galaxy NGC 4569 in Virgo, which is 
characterised by a truncated HI, CO, and dust disc ([\citenum{Boselli2006b}]) and by an extended tail of ionised gas ($\simeq$ 150 kpc in projected distance; [\citenum{Boselli2016a}]) 
witnessing an ongoing ram pressure stripping event. 

There is also indirect evidence that dust has been stripped in cluster galaxies along with the other ISM components. [\citenum{Longobardi2020a}] has observed a reddening of the UV colours 
of the Virgo cluster background galaxy population which they imputed as due to the differential light attenuation caused by the diffuse dust mixed within the ICM. Consistent results have been 
obtained in other clusters using similar studies based on the reddening of background galaxies ([\citenum{Chelouche2007}, \citenum{Muller2008}]), based on the far-infrared emission of the diffuse 
component associated with the ICM (e.g. [\citenum{Popescu2000}, \citenum{Giard2008}, \citenum{Planck2016a}, \citenum{Gutierrez2017}]), or using tuned simulations ([\citenum{Gjergo2018}, 
\citenum{Vogelsberger2019}]).

But so far the presence of dust associated to the stripped gas at large distances from the galaxy disc is still missing. The presence of dust in these stripped structures is fundamental for 
several reasons. First of all, this dust might be a major contributor to the pollution of the ICM, and thus be an alternative origin of the often claimed ejecta of supernovae in the early 
phase formation of the giant ellipticals inhabiting high-density regions ([\citenum{Renzini1993}, \citenum{Werner2008}, \citenum{dePlaa2013}, \citenum{Mernier2017}]). Its presence is also 
fundamental for understanding the fate of the stripped material since dust can contribute with metal lines to the cooling of the stripped gas, necessary for the formation of GMC where star 
formation can take place. Theoretical models of the ISM of galaxies suggest that dust cooling dominates on metal line cooling in low column density environments (e.g. [\citenum{Wolfire1995}]), 
in particular whenever the metallicity of the gas is low (e.g. [\citenum{Yamasawa2011}]). The physical condition of the gas in the stripped tails, where heating is probably dominated by 
the X-ray radiation of the surrounding medium, are significantly different from those encountered within the disc of star-forming galaxies. It is thus hard to identify which is the 
dominant cooling component, but it is likely that both metal lines and dust can contribute to this important physical process. Interesting is the fact that star formation occurs in most 
(e.g. [\citenum{Fossati2016}, \citenum{Fossati2018}, \citenum{Poggianti2019b}, \citenum{Gullieuszik2020}, \citenum{Boselli2023}]), but not all (e.g. [\citenum{Boissier2012}, 
\citenum{Boselli2016a}]), the stripped tails of cluster galaxies. If present, dust would play a major role in the energetic equilibrium of the cold stripped gas once mixed with the surrounding 
hot ICM thanks to its cooling properties. Its observation is thus fundamental for understanding the physics of the different components in this extreme environment and thus posing strong 
observational constraints to models and simulations. 

Only a couple of instruments are at present available to the community for the observation and the detection of dust in stripped tails. The MIRI instrument on the James Webb Space Telescope 
(JWST) can be used to detect dust emitting in the mid-IR spectral domain (5-27 $\mu$m; [\citenum{Bouchet2015}]). At these frequencies, however, the emission is dominated by the hot dust 
component which is probably only a very small fraction of the total dust stripped during the interaction. The NIRCam instrument ([\citenum{Rieke2023}]), which works in the 0.6-5 $\mu$m 
spectral domain, can be used to identify cold dust in absorption but only close to the galaxy disc where the stellar continuum emission is sufficiently bright, as already done using HST (see above). 
The Atacama Large Millimeter Array (ALMA) can be used to observe the very cold gas components emitting at wavelengths $\lambda$ $\geq$ 350-450 $\mu$m. Thanks to its excellent angular 
resolution, this radio array would be ideally suited to detect dust in clumpy structures where star formation takes place. Despite its extraordinary performances, however, the first 
attempts to detect cold dust in emission in stripped tails failed (Boselli et al., in prep.) probably because of the strong decrease of the spectral energy distribution of the emitting 
dust at mm-submm frequencies. Furthermore, ALMA does not cover the most interesting far-IR emission lines such as [CII] at 158$\mu$m which play a major role in the cooling process of the 
stripped gas.

The PRIMA mission (Glenn et al. 2025, this issue) will offer us a unique opportunity for the direct observation of dust in emission within the tails of local and distant stripped galaxies in 
dense environments. With its exceptional sensitivity in imaging and spectroscopic observing mode, we will have the opportunity to detect dust in a large range of peculiar environments, measure 
the gas-to-dust ratio, and finally understand whether stripping in dense environments might contribute to the pollution of the ICM. The polarimetric mode of PRIMA will also provide us with 
important measurements of the turbulent magnetic field within the tails, a parameter playing a fundamental role in confining the gas favoring its collapse into giant molecular clouds (GMC) 
where star formation can occur. This publication is intended to study the possibility of observing and detecting dust in emission with PRIMAger (Ciesla et al. 2025, this issue) and 
FIRESS Bradford et al. 2025, this issue) within the tails of some representative objects in local clusters. We already presented this idea in the dedicated PRIMA GO science white book 
published in 2022\footnote{https://arxiv.org/pdf/2310.20572}. We develop this idea here with more accurate estimates of the exposure times needed to get useful data for this proposed science case. 
The interested readers can find a detailed description of the PRIMA mission and of the performances of the two instruments on this dedicated JATIS issue at the references given above.

\section{Methodology}

\subsection{Dust column density}

The quantity of dust in the tails of stripped galaxies can be derived after measuring the gas column density in these regions using HI data at 21 cm and making some simple assumptions on 
the expected gas-to-dust ratio. Tails in HI without any associated stellar component, witnessing and ongoing ram pressure stripping process, have been detected in several objects in local 
clusters such as Virgo (e.g. [\citenum{Vollmer2003}, \citenum{Chung2007}, \citenum{Sorgho2017}, \citenum{Boselli2023}]), Coma ([\citenum{Gavazzi1989}, \citenum{Bravo-Alfaro2000}]), A1367 
([\citenum{Dickey1991},  \citenum{Scott2012}, \citenum{Scott2022}]), and Fornax ([\citenum{Lee-Waddell2018}, \citenum{Loni2021}, \citenum{Serra2023}, \citenum{Serra2024}]). They typically 
extend for several tens of kpc (projected distance) out from the stellar disc of the perturbed galaxies. Most of them have been formed after the interaction of galaxies with the surrounding 
ICM (ram pressure stripping), but there are also clear examples where the gas has been removed after a flyby gravitational encounter with other cluster members (e.g. NGC 4254, 
[\citenum{Vollmer2005}, \citenum{Haynes2007}, \citenum{Duc2008}, \citenum{Boselli2018b}]). There are also more extreme examples where atomic hydrogen clouds have been detected far 
from any known galaxy, where the gas is probably the remnant of an older perturbation ([\citenum{Kent2007, Kent2009, Haynes2007}]). These tails are all characterised by a gas column density 
of the order of $n(HI)$ $\simeq$ 1-10 $\times$ 10$^{19}$ cm$^{-2}$, or equivalently $n(HI)$ $\simeq$ 0.1-1 M$_{\odot}$ pc$^{-2}$ (e.g. [\citenum{Chung2007}]). 

In any kind of interaction, the gas component the most easily removed is the cold HI located in the outer disc and thus loosely bound to the gravitational potential well of the parent galaxy. 
Indeed, this gas component extends in unperturbed systems up to $\sim$ 1.8 the optical disc (e.g. [\citenum{Cayatte1994}]). Given the well known metallicity gradients in spiral galaxies, this 
gas is probably metal-poor, but it still contains an important content of dust. In the outer disc of M31, for example, it has been estimated that the dust-to-gas ratio is roughly 1/3 solar 
(e.g. [\citenum{Cuillandre2001}]). Similar results have been obtained in the outer disc of other local galaxies (e.g. [\citenum{Popescu2003, Holwerda2005}]). It is thus conceivable that dust, 
along with gas, can be stripped during these kind of interactions.

A first estimate of the dust column density within the tails of stripped material can be done assuming a constant gas-to-dust ($G/D$) ratio. For simplicity we use here the typical value observed 
in the solar neighborhoods ($G/D$ $\simeq$ 100, [\citenum{Bohlin1978}]), which should be considered as lower limit. Given the relative low-metallicity environment in the tail, and the tight 
relation between $G/D$ and gas metallicity observed in galaxy discs ([\citenum{Remy-Ruyer2014}]), a gas-to-dust ratio of $G/D$=1000 should be considered as an upper limit.  
The results that we obtain here can be easily scaled using more realistic values if requested. If we consider a gas column density of $n(HI)$ = 0.1 M$_{\odot}$ pc$^{-2}$, and we assume that 
the dust is diffuse and distributed as the atomic hydrogen with a $G/D$ $\simeq$ 100, the dust column density is $\Sigma(dust)$ $\simeq$ 10$^{-3}$ M$_{\odot}$ pc$^{-2}$. In local cluster 
galaxies the tails of stripped gas extend several tens of kpc in projected distance. If we consider as an example the perturbed galaxy NGC 4424 in the Virgo cluster, its HI tail as detected 
by recent MeerKAT observations gathered during the ViCTORIA survey ([\citenum{deGasperin2025}]) extends over an area $A$ $\simeq$ 100 kpc $\times$ 10 kpc and has a typical column density of 
$n(HI)$ = 0.1 M$_{\odot}$ pc$^{-2}$ (Fig. 1; see also [\citenum{Chung2007, Sorgho2017}]). The HI gas mass in the tail is $M_{HI}$ $\simeq$ 7.5 $\times$ 10$^7$ M$_{\odot}$, and thus the one 
expecterd for the cold dust is $M_{Dust}$ $\simeq$ 7.5 $\times$ 10$^5$ M$_{\odot}$.

\begin{figure}
\begin{center}
\begin{tabular}{c}
\includegraphics[height=15.5cm]{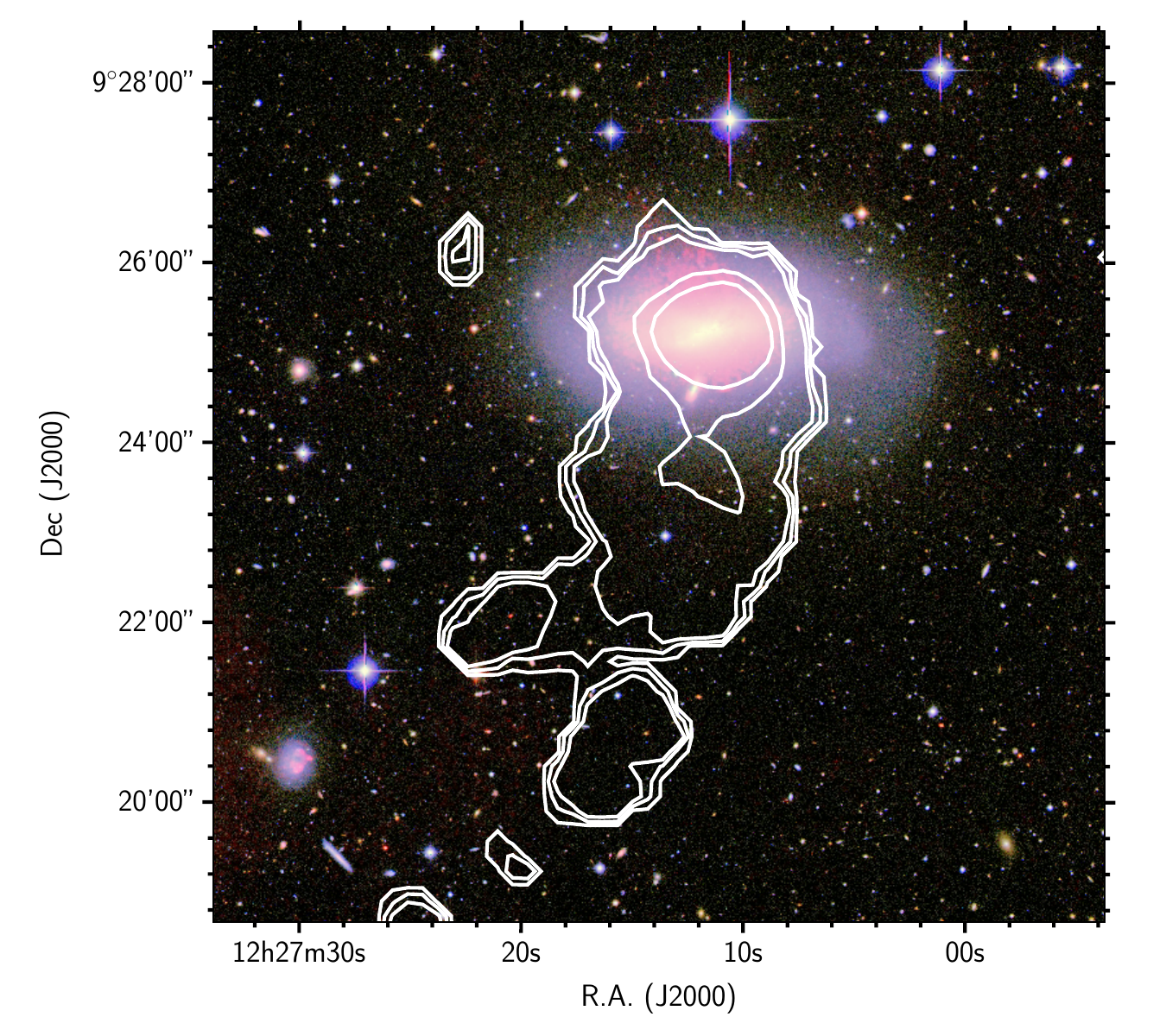}
\end{tabular}
\end{center}
\caption 
{ \label{N4424}
Pseudo-colour image of NGC 4424 in the Virgo cluster (1 arcmin = 4.8 kpc), adapted from [\citenum{Boselli2018c}]). The red diffuse low surface brightness feature extending in the north 
direction is ionised gas detected in H$\alpha$ during the VESTIGE survey ([\citenum{Boselli2018a}]). White contours show the HI gas distribution starting at column density 
$n(HI)$=10$^{19}$, 5$\times$10$^{19}$, 10$^{20}$, 5$\times$10$^{20}$, 10$^{21}$ cm$^{-2}$ (from [\citenum{Chung2009}]). For a $G/D$ ratio of $G/D$=100, which roughly corresponds 
to the expected upper limit in the $G/D$ ratio observed in ram pressure stripped tails, these contours correspond to $\Sigma_{dust}$ = 0.8, 4, 8, 40, 80$\times$10$^{-3}$ 
M$_{\odot}$ pc$^{-2}$, or at a distance of 16.5 Mpc to a surface brightness at 250 $\mu$m $\Sigma_{250}$ $\simeq$ 100, 500, 1000, 5000, 10000 kJy sr$^{-1}$.
These dust column densities and 250 $\mu$m surface brightnesses would drop by a factor of 10 for a $G/D$=1000. Realistic estimates should be in between these two limits.}
\end{figure}

\subsection{Expected far-infrared surface brightness}

We can now deduce the expected far-infrared flux density at different frequencies assuming that the dust spectral energy distribution (SED) is well approximated by a simple modified 
black-body as indicated by the relation:

\begin{equation}
S_{\nu} = \frac{M_{dust}}{D^2}k_{\nu}(\frac{\nu}{\nu_0})^\beta B(\nu,t)
\end{equation}

\noindent 
where $S_{\nu}$ is the far-infrared flux density at the frequency $\nu$, $D$ is the distance of the galaxy, $k_{\nu}$ the grain opacity at the frequency $\nu$ (also called absorption 
cross-section per mass of dust, measured in cm$^2$ g$^{-1}$), $\beta$ the variation as a function of frequency, and $B(\nu,T)$ a black body emission with a temperature $T$. The far-infrared 
flux density can thus be derived making some assumptions on the $k_{\nu}$, $T$, and $\beta$. For reference, assuming a mean dust temperature $T$ $\simeq$ 20 K, a grain opacity 
$k_{250}$ = 4 cm$^2$ g$^{-1}$, and 
$\beta$=0, this yields a total flux density over an area $A$ $\simeq$ 100 kpc $\times$ 10 kpc of 80 mJy, or equivalently a 250 $\mu$m surface brightness of $\Sigma_{250}$ $\simeq$ 22 
kJy sr$^{-1}$. It is hard to predict how the free parameters of Eq. (1) vary in the stripped gas of perturbed galaxies, where the physical conditions of the medium are still not fully 
constrained. The grain opacity parameter is expected to vary as a function of metallicity (e.g. [\citenum{Draine2007, Galliano2011}]) and density of the ISM (e.g. [\citenum{Kohler2015}]; 
for a recent review on dust properties we refer the reader to [\citenum{Galliano2018}]). In the Milky Way it has values ranging 3.7 
$\lesssim$ $k_{250}$ $\lesssim$ 6.4 cm$^{2}$ g$^{-1}$ ([\citenum{Bianchi2019}]). In the Large Magellanic Cloud, where the metallicity is $Z$ $\simeq$ 0.5$Z_{\odot}$ ([\citenum{Russell1992}], 
corrected by [\citenum{Asplund2009}]), the grain opacity at 160 $\mu$m is $k_{160}$ = 1.6 cm$^{2}$ g$^{-1}$ ([\citenum{Galliano2011}]), thus a factor of $\sim$ 1.6 larger than in the Milky 
Way ([\citenum{Draine2007}]). Stripping will preferentially remove the gas located in the outer disc of galaxies where it is less bound to the gravitational potential well of the disc 
(e.g. [\citenum{Domainko2006, Boselli2022}]). Given the well-known metallicity gradients in galaxies (e.g. [\citenum{Zaritsky1994}]), we thus expect that the gas in the tails is metal-poor, 
as indeed observed in some representative objects (e.g. [\citenum{Yoshida2012}] in the Coma cluster; $Z$ $\simeq$ 0.7$Z_{\odot}$ in ESO137-001, [\citenum{Fossati2016}]). On the other hand, 
the gas in these extreme environments has generally a low density, and might thus be characterised by a low grain opacity (e.g. [\citenum{Kohler2015}]). Even less constrained is the dust 
temperature, which has been only tentatively detected in emission in a few Virgo cluster galaxies by [\citenum{Longobardi2020b}] but at only one wavelength ($\lambda$ = 250$\mu$m), thus not 
allowing any direct estimate of this parameter. The physical conditions within the stripped gas are certainly different than those observed within the disc of unperturbed spiral galaxies. 
Indeed, there are indications that here the main heating source is not necessarily the interstellar radiation field since many tails of stripped gas do not host star forming regions. Several 
mechanisms have been invoked so far, and they include shocks ([\citenum{Nulsen1982, Roediger2006, Tonnesen2011}]), heat conduction and mixing ([\citenum{Cowie1977}]), magneto-hydrodynamic 
waves ([\citenum{Tonnesen2014, Ruszkowski2014}]). There is also evidence that the gas is stripped principally as cold atomic hydrogen, and once in the tail it gets first ionised, than hot gas 
emitting in X-rays ([\citenum{Boselli2016a, Boselli2021}]). It is thus hard to infer any possible range in the expected dust temperature. In the lack of any accurate constrain, we can assume 
that the dust temperature is similar to the one observed in local, star forming galaxies (15 $\lesssim$ $T$ $\lesssim$ 30 K). Finally, we can leave $\beta$ as a free parameter, remembering 
that in normal galaxies it ranges in between 1$\lesssim$ $\beta$ $\lesssim$ 2.5, with low values in low metallicity environments 
([\citenum{Galliano2005, Bendo2006, Galametz2010,  O'Halloran2010, Boselli2012}]). 

\begin{figure}
\begin{center}
\begin{tabular}{c}
\includegraphics[height=15.5cm]{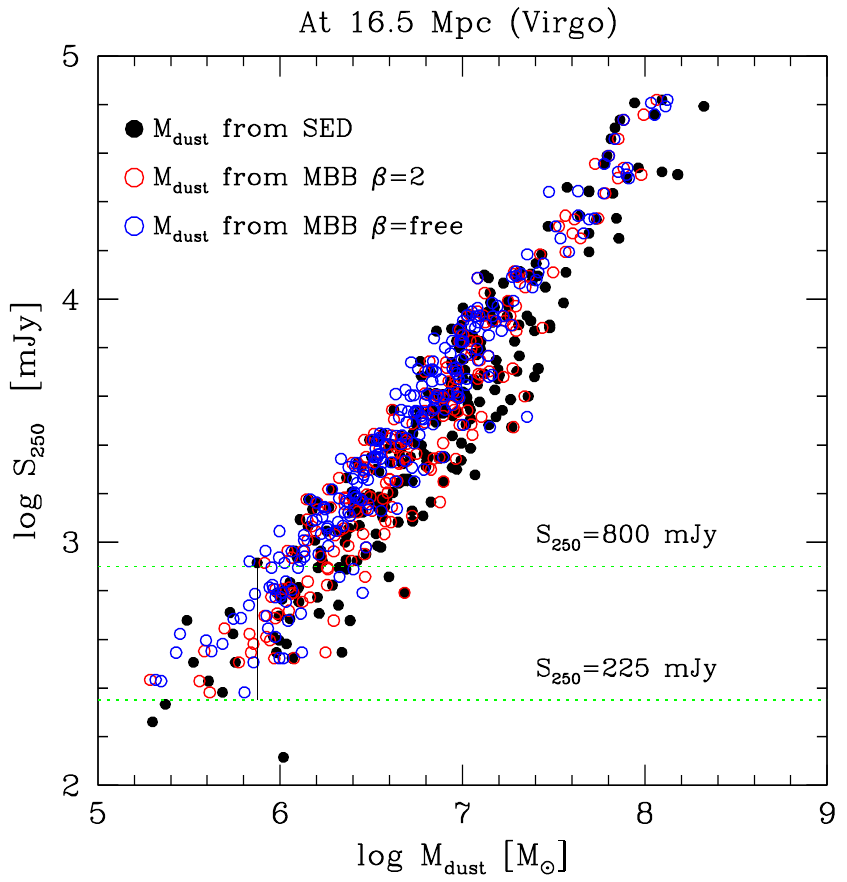}
\end{tabular}
\end{center}
\caption 
{ \label{massrange}
Relationship between the far-infrared flux density $S_{250}$ measured at 250 $\mu$m (from [\citenum{Ciesla2012}]) and the dust mass of the HRS galaxies where $M_{dust}$ is measured using a 
modified black body with $\beta$=2 (red open circles) or $\beta$ as a free parameter (blue open circles; both from [\citenum{Cortese2014}]), or using a SED fitting analysis using the 
[\citenum{Draine2007}] models, from [\citenum{Ciesla2014}]; black filled dots). For a fair comparison, the relation has been derived after scaling all galaxies at the distance of the Virgo 
cluster (16.5 Mpc). The vertical black solid line show the dynamic range in $S_{250}$ (in mJy) expected for a tail of dust mass $M_{dust}$ = 7.5$\times$10$^5$ M$_{\odot}$ derived for a $G/D$=100 
as the one expected in NGC 4424 (Fig. \ref{N4424}). Under these conditions, the expected integrated flux density in the tail ranges in between 225$\lesssim$ $S_{250}$ $\lesssim$ 800 mJy.} 
\end{figure}

These are standard conditions for the dust properties in the ISM of local star-forming galaxies. We can thus use their observed quantities to derive the expected far-infared surface brightness 
of the dust in a typical tail. For this purpose we use the dust masses derived for the HRS, a complete $K$-band selected sample of local galaxies (15$\leq$ $D$ $\leq$ 25 Mpc) spanning a wide 
range in morphological type, stellar mass, star formation rate, and metallicity ([\citenum{Boselli2010}]). For this sample, which is composed by 260 late-type galaxies including cluster 
perturbed systems such as those analysed in this work, three different and independent set of dust mass estimates are available in the literature. Two comes from [\citenum{Cortese2014}], 
where dust masses are derived by fitting the far-infrared SED in the 100-500$\mu$m spectral domain using a modified black body with $\beta$=2, or with $\beta$ as a free parameter. A third 
dust estimate has been measured by fitting the observed near- to far-infrared SED using the CIGALE fitting code ([\citenum{Boquien2019}]) using the [\citenum{Draine2007}] models 
([\citenum{Ciesla2014}]). Figure \ref{massrange} shows the relationship between the far-infrared flux density $S_{250}$ measured at 250 $\mu$m (from [\citenum{Ciesla2012}]) and the dust 
mass where $M_{dust}$ is derived using the three different techniques mentioned above. For comparison with galaxies in the Virgo cluster, the relation is derived after scaling the dust 
masses at the distance of 16.5 Mpc ([\citenum{Gavazzi1999, Mei2007, Cantiello2024}]).

Figure \ref{massrange} shows that for a dust mass of $M_{dust}$ $\sim$ 7.5$\times$10$^5$ M$_{\odot}$ derived assuming a $G/D$ = 100, the expected far infrared flux density at 250 $\mu$m is within 
the range 225$\lesssim$ $S_{250}$ $\lesssim$ 800 mJy for a tail extended over 100$\times$10 kpc. The corresponding surface brightness is 61$\lesssim$ $\Sigma_{250}$$\lesssim$610 kJy sr$^{-1}$.

\subsection{PRIMAger observations}

The far-infared diffuse emission of the dust in the tail can be detected in continuum using the PRIMAger camera. We can use the sensitivity calculator to estimate the exposure time necessary to 
map an area of 100kpc $\times$ 10 kpc (which at the distance of the Virgo cluster corresponds to 1250 arcsec$\times$125 arcsec) to detect the diffuse dust emission at a given signal-to-noise ($S/N$)
ratio. Given the extension of the tail on the sky, observations should be undertaken using the stirring mirror. We recall that this area corresponds to one of the most extended ram pressure gas tail 
ever observed in the local Universe. This number should thus be considered as an upper limit for this exercise. 

Here we simulate the observations for a 250$\mu$m flux density of $S_{250}$ $\simeq$ 350 mJy, corresponding on this area to a surface brightness $\Sigma_{250}$ $\simeq$ 100 kJy sr$^{-1}$. 
This number corresponds to a low value derived from Fig. \ref{massrange} for a $G/D$ ratio $G/D$ = 100 typical for a gas with a solar metallicity. These estimates can be easily scaled to 
different, more representative values if necessary. For reference we also calculate the integration times necessary to detect the dust emission if this is measured assuming a more extreme $G/D$ = 1000, 
in agreement with what recently deduced by [\citenum{Longobardi2020b}] in the tails of a few ram pressure stripped galaxies in Virgo. We recall that a gas-to-dust ratio $\simeq$ 1000 is observed in very 
low metallicity environments ($Z$ $\simeq$ 1/10 $Z_{\odot}$) such as those of dwarf star-forming systems ([\citenum{Remy-Ruyer2014}]), where the metallicity is probably lower than the one 
observed in the tails of stripped massive galaxies (e.g., [\citenum{Fossati2016}]).

According to the PRIMAger expected sensitivity reported in Ciesla et al. (this issue), a surface brightness of $\Sigma_{235}$ $\simeq$ 180 kJy sr$^{-1}$ (5$\sigma$) can be reached in 
10 hours of integration over 1 square degree (including overheads).
Scaling it to the area covered by the tail ($\simeq$ 43 arcmin$^2$), and to the expected surface brightness estimated above, this leads to 23 min of integration to detect the continuum 
dust emission with a signal-to-noise S/N=5. This number can be significantly reduced if mapping is limited to selected regions or reducing the signal-to-noise to S/N=1 ($\sim$ 1 min), which 
is sufficient considering that the signal is spatially correlated on scales of tens of arcminutes and is thus well resolved in the imaging data.
Adopting a more extreme $G/D$ $\simeq$ 1000, a surface brightness of $\Sigma_{235}$ $\simeq$ 10 kJy sr$^{-1}$ could be reached in $\sim$ 38h of integration time at a S/N=5 and $\sim$ 1.5h at a S/N=1. 
The emission of the tail will be identified and recognised from that of the confusion-limited background, which is expected to be at 235 $\mu$m of $\sim$ 15 mJy, [\citenum{Donnelan2024}], 
since correlated on large scales and with an expected distribution similar to the one observed in the different gas phases. We can thus conclude that, since the $G/D$ ratio in the tails of 
ram pressure stripped galaxies is expected to be in the range 100 $\lesssim$ $G/D$ $\lesssim$ 1000 given the observed gas metallicity (e.g. [\citenum{Fossati2016}]), dust can be detected with PRIMAger.  
If the dust is detected at different frequencies so that a SED can be derived, the data can also be used to infer a typical dust temperature, a critical parameter necessary to constrain models 
and simulations with the aim of understanding the gas phase transition within the tails. 

PRIMAger offers also the possibility to undertake polarimetry observations,
which are of great interest since they can be used to quantify the strength of the turbulent magnetic field ([\citenum{Andersson2015}]). These measurements can be compared to those that SKA 
observations will provide for the ordered magnetic field in the radio domain. Recent hydrodynamic simulations have indicated the magnetic field as a crucial parameter in confining the stripped 
material, thus self-shielding it from the heating due to the surrounding ICM, triggering gas collapse and star formation ([\citenum{Tonnesen2014, Ruszkowski2014}]). The degree of polarisation 
observed in radio continuum in similar tails of stripped material is still poorly known since observed in a very few objects. It ranges between 5\%\ and 30\%\ (NGC 4522, [\citenum{Vollmer2004}]). 
We do not have so far any direct estimate of the degree of polarisation of the dust within these tails of stripped material. We can thus just make some rough estimates. The dominant component 
of the ISM removed in any kind of interaction is the cold atomic gas phase since it is the one less bound to the gravitational potential well of the perturbed galaxy ([\citenum{Boselli2022}]), 
as indeed shown in Fig. \ref{N4424}. The typical column density of the gas in these tails is generally in the range 10$^{17}$ $\lesssim$ $n(HI)$ $\lesssim$ 10$^{21}$ cm$^{-2}$, with regions 
of low column density dominant in space. High column densities are observed only in clumpy regions, where star formation can take place. We recall, however, that star formation in stripped 
tails is not ubiquitous ([\citenum{Boselli2022}]). The Planck survey of the Milky Way has shown that at 353 GHz (850 $\mu$m) the linear polarisation fraction is of the order of $\sim$ 20\%\ , 
and that this number decreases with increasing gas column density in the range 10$^{20}$ $\lesssim$ $n(HI)$ $\lesssim$ 10$^{22}$ cm$^{-2}$ ([\citenum{Planck2020}]). At these and at shorter 
wavelengths the polarisation is due to the thermal emission of dust, and does not seem to strongly depend on the dust temperature ([\citenum{Planck2020}]). The physical conditions of the 
ISM in the stripped tails, where the gas is principally heated by different processes (heat conduction, mixing, magnetohydrodynamic waves, shocks), are very different than in the Milky Way 
(photoionisation), it is thus very difficult to predict which is the degree of thermal polarisation of dust grains. Given the density of the gas in the tails (see above), we can imagine that 
polarisation is comparable to the one observed in the diffuse medium of the Milky Way. It might probably drop in the densest regions where star formation takes place, as indeed observed in 
some star forming regions in the Milky Way with Planck (polarisation fraction of $\sim$ 4\%\ in the Rosetta Nebula, [\citenum{Planck2016b}]). Here, however, the far-IR emission of the dust 
is expected to be higher given the high gas column density of the gas.

Assuming a polarisation of $\sim$ 10\%\ we can calculate the integration time necessary to measure it with the same S/N=1 than in imaging mode. This can be done 
in $\sim$3h of integration for a $G/D$ = 100, while the integration times becomes prohibitive if $G/D$ = 1000. We recall, however, that these integration times can be significantly reduced 
rebinning the data on extended regions (Voronoi binning), which is possible thanks to the very extended nature of these tails, or reducing the sky mapping on a few representative regions 
within the tail. 

\subsection{FIRESS observations}

The spectrograph FIRESS on PRIMA also offers a unique opportunity for the study of the dust and gas properties within the tails of stripped material in cluster galaxies. It can be used to 
measure several infrared line emissions and thus to constrain the composition, density, and temperature of the stripped ISM. In the spectral domain covered by FIRESS (24-240 $\mu$m) there 
are several emission lines of primordial importance for constraining the properties of the stripped gas. The brightest of them is the [CII] line at 158 $\mu$m, which is one of the principal 
coolers of the ISM in star forming galaxies (\citenum{Croxall2017}]). Other lines (e.g. [NII] 122 $\mu$m and [NII] 205 $\mu$m), characterised by different ionising potentials and 
critical densities for collisions with electrons, can be used to identify the origin of the [CII] emission (ionised gas, molecular gas, atomic gas) in the stripped material, the 
electron density and the metallicity of the stripped gas, or the photoelectric heating efficiency using the [CII]158 $\mu$m+[OI]63 $\mu$m/TIR ratio ([\citenum{Croxall2017, Kewley2019}]). 
Variations of the dust column density and temperature are critical to measure the timescales for dust sputtering once stripped from the cold galaxy ISM and mixed within the hot ICM. 

Targeted far-IR spectroscopic observations of stripped tails in cluster galaxies have not been undertaken so far. A few Virgo cluster galaxies, some of which hosting tails of stripped 
material, have been observed with the Long Wavelength Spectrograph (LWS) onboard of the Infrared Space Observatory (ISO) ([\citenum{Leech1999}) and with the Fourier Transform Spectrograph 
(FTS) of the SPIRE instrument onboard of Herschel ([\citenum{Minchin2022}]). Because of the limiting sensitivity of these observations ($\sim$ 10$^{-9}$ and 10$^{-10}$ W m$^{-2}$ sr$^{-1}$, 
respectively), none of these tails has ever been detected.
As for the dust continuum emission, we have to rely to first order estimates. Quantifying the expected contribution of these different cooling lines ([CII]158 $\mu$m, [OI]63 $\mu$m, 
[NII]122 $\mu$m and 205 
$\mu$m) is particularly difficult given that the physical condition of the stripped gas are very different than those encountered in other galacitic environments where dust heating is 
principally due to the interstellar radiation field. We can thus make only a very speculative estimate making the strong assumption that their emission properties are similar to those 
observed in the ISM of star forming systems.
Assuming the typical line-to-total infrared emission (TIR) ratios of [\citenum{Croxall2012, Croxall2017}] ([CII]158 $\mu$m/TIR $\sim$ 0.005, 
[CII]158 $\mu$m/[OI]63 $\mu$m $\sim$ 5, [CII]158 $\mu$m/[NII]205 $\sim$ 10, and [NII]122/205 $\mu$m $\sim$ 1-8) and the $S_{250}$ vs. $S_{TIR}$ ratio of [\citenum{Galametz2013}] 
($\nu$$S_{250}$ $\sim$ 0.3 $S_{TIR}$) observed in star-forming galaxies, their expected surface 
brightness in the diffuse medium should be $\Sigma_{[CII]158}$ $\simeq$ 
2$\times$10$^{-12}$ W m$^{-2}$ sr$^{-1}$, 
$\Sigma_{[OI]63}$ $\simeq$ 
4$\times$10$^{-13}$ W m$^{-2}$ sr$^{-1}$, 
 $\Sigma_{[NII]205}$ $\simeq$ 2$\times$10$^{-13}$ W m$^{-2}$ sr$^{-1}$, 
and $\Sigma_{[NII]122}$ $\simeq$ 2$\times$10$^{-13}$ W m$^{-2}$ sr$^{-1}$ (see Table \ref{tablines}). 
We can estimate the integration times necessary to detect the selected lines in the diffuse medium by scaling the nominal FIRESS sensitivity of 3$\times$10$^{-19}$ W m$^{-2}$ (5$\sigma$) 
reached over 100 arcmin$^2$ in line emission in 800 hours in the spectral range 24$\leq$ $\lambda$ $\leq$ 75 $\mu$m and in 336 hours $\times$ ($\lambda$/100 $\mu$m)$^{-1.68}$ in the range 
75 $<$ $\lambda$ $\leq$ 235 $\mu$m to the surface brightnesses given above. The integration times necessary to map an area of 43.4 arcmin$^2$ for a 1$\sigma$ detection are given in 
Table \ref{tablines}. The [CII]158$\mu$m line can be achieved in $\sim$ 8.5 h over the full sampled region, while the detection of the remaining lines is prohibitive at 1 arcmin resolution.
These integration times can be significantly reduced in selected regions where star formation takes place since here the density of the gas is expected to increase by at least a factor 
of 40 (see next Sec.) with respect to the one adopted for the calculations reported above. We should, however, remember that these star-forming regions have sizes of a few hundreds 
parsec, and might not be fully resolved even in very nearby systems. 
Overall, with the diffuse emission potentially extended to $>$1000 beams across, FIRESS integration times can be significantly reduced using Voronoi binning on limited regions to under 
20h for each line and spectral setting, making these important observations accessible at least on a selected number of targets. 

\begin{table}[ht]
\caption{FIRESS estimated integration times} 
\label{tablines}
\begin{center} 
\begin{tabular}{|l|c|c|} 
\hline
\rule[-1ex]{0pt}{3.5ex}  Line           & Surf. Brightness      & Int. time \\
\hline
\rule[-1ex]{0pt}{3.5ex}  Units          & W m$^{-2}$ sr$^{-1}$  & h         \\ 
\hline\hline
\rule[-1ex]{0pt}{3.5ex}  [OI]63$\mu$m   & 4$\times$10$^{-13}$   & 1089  \\ 
\hline
\rule[-1ex]{0pt}{3.5ex}  [NII]122$\mu$m & 2$\times$10$^{-13}$   & 1310  \\
\hline 
\rule[-1ex]{0pt}{3.5ex}  [CII]158$\mu$m & 2$\times$10$^{-12}$   & 8.5   \\
\hline
\rule[-1ex]{0pt}{3.5ex}  [NII]205$\mu$m & 2$\times$10$^{-13}$   & 548   \\
\hline 
\end{tabular}
\end{center}
Note: measured to reach 1$\sigma$ detection over 43.4 arcmin$^2$ at a resolution of 1 arcmin. Integration times scales as sensitivity$^2$ $\times$ FoV $\times$ resolution$^{-2}$. 
The numbers reported here have been derived for a gas column density of $n(HI)$ = 0.1 M$_{\odot}$ pc$^{-2}$, the typical column density of the diffuse component. 
All estimates assume $G/D$=100.
\end{table} 

\subsection{Target selection}

The science case presented above and the estimates of the integration time necessary to detect the tail of stripped objects both with PRIMAger and FIRESS have been tuned for the 
observation of a representative galaxy in the Virgo cluster (NGC 4424, Fig. \ref{N4424}). Objects with similar properties, however, have been detected in many other massive clusters 
and groups (e.g. [\citenum{Boselli2022}]) in the local Universe and up to redshift $z$=0.7 ([\citenum{Boselli2019}]). Being the closest cluster of galaxies, Virgo has been chosen for this 
exercise since it allows us to reach the best angular resolution of the tails, and thus potentially resolve any possible substructure within them. At the distance of the cluster (16.5 Mpc, 
[\citenum{Gavazzi1999, Mei2007, Cantiello2024}]) 1 arcsec = 80 pc, thus the resolution of the two instruments corresponds to $\sim$ 0.3-2.2 kpc. We recall that the typical size of the 
star-forming complexes observed within the tails of stripped galaxies in Virgo are of the order of a few hundred parsecs, the thickness of the filaments of atomic or ionised gas up to a 
few kpc, while their length can exceed a hundred kpc in projected distance. Integration times can be significantly reduced by selecting farther targets where the tails have a smaller angular 
dimension, or more extreme objects characterised by large quantities of molecular gas ([\citenum{Moretti2018, Moretti2020}]) and massive star clusters in the tail, as those identified 
during the GHASP survey ([\citenum{Poggianti2017}]), where the detection of the stripped dust is potentially less challenging, but to the detriment of angular resolution. Furthermore, 
the identification of perturbed galaxies in Virgo has been done through the analysis of multifrequency blind surveys spanning the whole electromagnetic spectrum, from the UV 
(\citenum{Boselli2011}), to the visible ([\citenum{Ferrarese2012}]), H$\alpha$ narrow-band imaging ([\citenum{Boselli2018a}]), far-IR ([\citenum{Davies2010}]), HI and radio 
continuum ([\citenum{deGasperin2025}]). The sample of perturbed galaxies identified during these surveys is complete and thus less biased than other samples which might be strongly 
dominated be extreme objects. A combination of various targets extracted from different samples would probably be the ideal strategy for a key observing program.

\section{Discussion}

The analysis presented in the previous sections suggests that the observations of the dust component located in the tails of ram pressure stripped galaxies is probably reachable with 
both instruments on PRIMA, although challanging with FIRESS, also considering that the sensitivities used for these calculations are the minimal requested sensitivities for the two 
instruments, and could thus be overtaken in the next future. The estimates obtained in the previous section, however, are subject to large uncertainties since they depend on several 
unconstrained parameters. We should first remember that the gas-to-dust ratio in these extreme environments might be significantly different than in the ISM of the disc of spiral galaxies 
here used to derive the far-infrared emission in the tails. Indeed, the analysis of [\citenum{Longobardi2020b}] suggests that in these extreme environments the gas-to-dust ratio can be 
$G/D$ $\sim$ 1000, thus a factor of $\sim$ 10 higher than in the solar neighborhood. We recall, however, that this estimate suffers from the same uncertainties of our derivation since 
in [\citenum{Longobardi2020b}] dust masses have been measured using a single far-infrared band ($S_{250}$), making similar assumptions on the grain opacity, $\beta$, and dust temperature, 
all unconstrained parameters. The only parameter that can be measured on the available data is the gas metallicity. Direct measurements undertaken with MUSE suggest that the metallicity 
in the tail of stripped galaxies is slightly under solar (e.g. $Z$ $\sim$ 0.7 $Z_{\odot}$, [\citenum{Fossati2016}]). Given that extended tails of stripped material are mainly observed 
in massive systems, we can  expect that here the gas-to-dust ratio is within the range adopted in our calculations (100 $\lesssim$ $G/D$ $\lesssim$ 1000).

It is conceivable that all the other parameters characterising dust properties are different than in the ISM, where the heating source of the dust is mainly the interstellar 
radiation field. The estimates given above should thus be considered with caution. They should be representative for the diffuse medium where the gas has a relatively low column density. 
As mentioned above, however, in some tails star formation has been observed in relatively compact HII regions. This occurs only whenever the gas column density is higher than a typical 
threshold, which in tails of stripped gas embedded in a hot X-ray emitting gas is of the order of $n(HI)$ $\simeq$ 5$\times$10$^{20}$ cm$^{-2}$ (4 M$_{\odot}$ pc$^{-2}$) as derived from 
theoretical arguments ([\citenum{Taylor2005}]). These gas densities are a factor of $\sim$ 40 higher than those adopted for our estimates, it is thus conceivable that the detection of dust 
in continuum, polarised, and in several emission lines becomes significantly easier in clumpy regions where star formation takes place, provided that these regions are barely resolved during 
the observations. We recall, however, that the formation of new stars in the tails of stripped gas, if present in several objects, is not ubiquitous. There are, indeed, several examples 
where star formation is not present in the tail. The galaxy NGC 4424 shown in Fig. \ref{N4424} is a clear example. The deep H$\alpha$ image of the galaxy obtained during the VESTIGE 
project (A Virgo Environmental Survey Tracing Ionised Gas Emission, [\citenum{Boselli2018a}]) does not show any compact H$\alpha$ emission in the tail, nor clumps of young stars are 
detected in the deep broad-band images gathered during the Next Generation Virgo Survey (NGVS, \citenum{Ferrarese2012}]). Other clear examples are the bright spiral galaxy NGC 4569 in 
Virgo ([\citenum{Boselli2016a}]), a few other galaxies still in Virgo ([\citenum{Boissier2012}]), or the peculiar objects CGCG 97-73 and CGCG 97-79 in A1367 ([\citenum{Gavazzi2001, Pedrini2022}]). 
Furthermore, even within the tails hosting star forming regions, the principal heating source is stellar radiation only in limited regions close to the newly formed star clusters 
([\citenum{Fossati2016, Poggianti2019a}]). At larger distances other mechanisms can contribute to the heating of the dust and the ionisation of the gas (shock, heat conduction and mixing, 
magneto-hydrodynamic waves). Observations seem to indicate that the gas (and its associated dust) is principally stripped in its cold atomic phase, then it changes of phase becoming first 
ionised, than hot, once mixed with the hot surrounding ICM (e.g. [\citenum{Boselli2016a, Boselli2021, Boselli2022}]). It is known that dust grains are destroyed by thermal sputtering in 
the hot gas halo of elliptical galaxies ([\citenum{Bocchio2012}]). We might thus expect that the same happens once the stripped ISM heats up to $T$ $>$ 10$^7$ K as observed in some nearby 
objects (e.g. ESO 137-001, [\citenum{Sun2006}]).

There are, however, a few arguments suggesting that dust can be present and survive in the tails of some stripped galaxies. First of all, its presence has been deduced using the Balmer 
decrement in some star forming regions far from the galaxy discs (e.g. [\citenum{Fossati2016, Poggianti2019b}]). Furthermore, there are also indications of the presence of molecular gas 
in some of these tails ([\citenum{Jachym2014, Jachym2017, Jachym2019, Moretti2018, Moretti2020, Cramer2021}]), or predicted by simulations ([\citenum{Boselli2021}]). Although this gas, 
measured via the CO emission lines, is often located within limited regions where star formation takes place, its presence is generally associated to that of dust just because the molecular 
gas is efficiently formed on dust grains ([\citenum{Hollenbach1971, Wakelam2017}]). Finally, recent simulations suggest that in the presence of magnetic field the stripped material remains 
confined along filaments, where the density is sufficiently high to limit gas mixing and evaporation ([\citenum{Tonnesen2014, Ruszkowski2014}]). It is thus conceivable than in these 
regions the stripped dust is less subject to external events, and can thus survive sufficiently long to keep its typical properties observed within the ISM of unperturbed systems. 
Simulations indicate that the molecular gas phase can be present in the tails of stripped galaxies on timescales of a few hundreds Myr (e.g. [\citenum{Boselli2021}]), consistently with its 
observation at different distances from the galactic disc in some perturbed systems (e.g. [\citenum{Jachym2019}]). These timescales are slightly smaller than those necessary to produce 
the observed tails of stripped galaxies traveling at $\sim$ 1000 km s$^{-1}$ across the cluster ($\simeq$ 500 Myr). These timescales can be compared to those estimated for thermal dust 
sputtering destruction in a hot gas ($T$ $\geq$ 10$^6$ K). This can be derived following the formalism of [\citenum{Draine1979}] which gives timescales of the order of $\tau_{sput}$ $\simeq$ 10$^8$ 
yr for dust grains of size $a$ $\simeq$ 0.01 $\mu$m embedded in a hot gas of density $n(H)$ $\simeq$ 10$^{-4}$ cm$^{-3}$, or a factor of $\sim$ 10 shorter in small grains and PAHs 
([\citenum{Bocchio2012}]). 

All these arguments consistently suggest that, under some conditions, dust can survive in the hostile environment of stripped tails, thus becoming an excellent target for PRIMA observations. 
The data that this fabulous instrument will provide will be of primordial importance for constraining the physical properties of the ISM in these extreme environments, crucial for the definition 
of tuned models and simulations. 

\section{Conclusion}

The analysis presented in this work suggests that the PRIMAger camera and the FIRESS spectrograph on PRIMA are well suited to detect the dust frozen within the ISM and stripped in cluster 
galaxies during their interaction with the hostile surrounding environment. Although the observations remain challenging, they can be optimised by selecting ideal targets and representative 
regions within the tail to measure the dust associated far-infrared emission at different frequencies, the degree of polarisation, and several major emission lines in the far-infrared domain. 
All these 
data are of crucial importance for measuring the main parameters of the stripped dust and gas components, such as their temperature, density, emission properties, and identify the main heating 
source of the [CII] emission line (ionised, molecular, atomic gas). They can also be used to quantify the electron density and the metallicity of the stripped gas, along with the photoelectric 
heating efficiency, all fundamental parameters necessary to constrain models and simulations.

\section{Acknowledgments}

We warmly thank the two anonymous referees whose comments and suggestions helped improving the quality of the manuscript.

\section{Data availability}

The data presented in this article are publicly available on the
HeDaM database at the link: 

https://hedam.lam.fr/HRS/

\section{Disclosures}

The authors declare that there are no financial interests, commercial affiliations, or other potential conflicts of interest that could have influenced the objectivity of this research 
or the writing of this paper.

\bibliography{report} 
\bibliographystyle{spiebib} 

\end{spacing}
\end{document}